\documentclass[twocolumn,prd,preprintnumbers,amsmath,amssymb,nofootinbib]{revtex4}

\usepackage{graphicx}
\usepackage{dcolumn}
\usepackage{bm}
\usepackage{epsfig}
\usepackage{epstopdf}
\usepackage{amssymb}
\usepackage{amsmath}
\usepackage{subfigure}
\usepackage[utf8]{inputenc}

\newcommand{\beq}{\begin{equation}}
\newcommand{\eeq}{\end{equation}}
\newcommand{\Mpl}{M_\mathrm{Pl}}

\newcommand{\gtil}{\tilde{g}}

\newcommand{\TJi}{T_{J,i}}
\newcommand{\phimin}{\phi_\mathrm{min}}
\newcommand{\phimax}{\phi_\mathrm{max}}
\newcommand{\varphimax}{\varphi_\mathrm{max}}
\newcommand{\varphimin}{\varphi_\mathrm{min}}

\newcommand{\phidot}{\dot{\phi}}
\newcommand{\rhodot}{\dot{\rho}}
\newcommand{\phidotmax}{\phidot_\mathrm{max}}
\newcommand{\phibar}{\bar{\phi}}
\newcommand{\Vbar}{\bar{V}}
\newcommand{\Kbar}{\bar{K}}

\newcommand{\Pbar}{\bar{P}}
\newcommand{\rhobar}{\bar{\rho}}
\newcommand{\omk}{\omega_k}

\newcommand{\kphys}{k_\mathrm{phys}}
\newcommand{\phita}{\bar{\phi}_\mathrm{ta}}
\newcommand{\Veff}{V_\mathrm{eff}}
\newcommand{\Vmax}{V_\mathrm{max}}
\newcommand{\Kmax}{K_\mathrm{max}}
\newcommand{\delp}{\Delta p}
\newcommand{\phiprime}{\varphi^\prime}
\newcommand{\phiprimeavg}{\phiprime_\mathrm{avg}}
\newcommand{\phiprimemax}{\varphi'_\mathrm{max}}
\newcommand{\vel}{\dot{\phi}_M}

\newcommand{\mindot}{\dot{\phi}_\mathrm{min}}

\begin{document}

\title{Quartic Chameleons: Safely Scale-Free in the Early Universe}

\author{Carisa Miller}
\email{carisa@live.unc.edu}
\author{Adrienne L. Erickcek}
\email{erickcek@physics.unc.edu}
\affiliation{Department of Physics and Astronomy, University of North Carolina at Chapel Hill, Phillips Hall CB3255, Chapel Hill, North Carolina 27599, USA}


\begin{abstract}
In chameleon gravity, there exists a light scalar field that couples to the trace of the stress-energy tensor in such a way that its mass depends on the ambient matter density, and the field is screened in local, high-density environments.  Recently it was shown that, for the runaway potentials commonly considered in chameleon theories, the field's coupling to matter and the hierarchy of scales between Standard Model particles and the energy scale of such potentials result in catastrophic effects in the early Universe when these particles become nonrelativistic. Perturbations with trans-Planckian energies are excited, and the theory suffers a breakdown in calculability at the relatively low temperatures of Big Bang Nucleosynthesis. We consider a chameleon field in a quartic potential and show that the scale-free nature of this potential allows the chameleon to avoid many of the problems encountered by runaway potentials. Following inflation, the chameleon field oscillates around the minimum of its effective potential, and rapid changes in its effective mass excite perturbations via quantum particle production.  The quartic model, however, only generates high-energy perturbations at comparably high temperatures and is able remain a well-behaved effective field theory at nucleosynthesis.
\end{abstract}

\maketitle
\section{Introduction}
\label{sec:Intro}

Many explanations for the current accelerated expansion of the Universe posit the existence of a new light scalar field. These scalar fields are usually coupled to matter and so can mediate long-range forces, often of gravitational strength. Such scalars are not only cosmologically motivated, but also pervasive in high-energy physics and string theory. However, stringent experimental bounds imply tight constraints on any new fifth forces mediated by scalar fields. These constraints require the scalar's coupling to matter to be tuned to unnaturally small values in order to avoid detection. Another approach is to employ a screening mechanism, which suppresses effects of the field locally, allowing consistency with successful tests of general relativity.

One of the few known screening mechanisms capable of reconciling the predictions of scalar-tensor gravitational theories and experimental constraints is the chameleon mechanism \cite{KhouryShort, KhouryLong}. In chameleon gravity theories, the scalar field's potential function and its coupling to the stress-energy tensor combine into an effective potential whose minimum is dependent on the matter density of its environment. Consequently, the effective mass of the chameleon field is also dependent on the environment, increasing enough in regions of high density to suppress the field's ability to mediate a long-range force. Because of this ability to hide within its environment, the chameleon can couple to matter with gravitational strength and still evade experimental detection in laboratory and Solar System tests of gravity.

The vast majority of cosmological investigations of chameleon gravity have considered potentials of the runaway form, such as the exponential \mbox{$V(\phi)=M^4\exp[(M / \phi)^n]$} and power-law $V(\phi)=M^{4+n}\phi^{-n}$ potentials.  In order to evade Solar System tests of gravity, $M$ has to be set to a value of $\sim\!\!10^{-3}$ eV, which is the energy scale of dark energy \cite{KhouryLong}. This coincident energy scale gave the chameleon a lot of attention early on as a possible explanation for cosmic acceleration. However, it was shown in Ref. \cite{NoGo} that the chameleon field cannot account for the accelerated expansion of the Universe without including a constant term in its potential. Nevertheless, light scalar fields arise in many theories that consider physics beyond the Standard Model (SM), and the chameleon mechanism remains one of the most-studied approaches to screening the unwanted forces mediated by these fields.

Many laboratory experiments have been conducted to search for forces mediated by chameleon fields.  Experiments that use atom \cite{atom1} and neutron \cite{neutron1,neutron2} interferometry and those that use $\mu$m-sized test masses \cite{microshperes} have already placed constraints on chameleon theories. Additional experiments have been proposed: one aims to measure the interactions between parallel plates to search for new forces \cite{pp1} and another suggests using atom interferometry between parallel plates of different densities to detect density-dependent chameleon forces \cite{pp3}. Laboratory searches for chameleon particles converted from photons in the presence of a magnetic field via the Primakov effect have placed constraints on the chameleon-photon coupling \cite{ADMX,Chase}. The CERN Axion Solar Telescope searched for chameleons created in the Sun by this effect \cite{sun2} and is currently conducting more sensitive searches \cite{sun4} to detect solar chameleons via their radiation pressure \cite{sun1}. Chameleon theories have also been constrained by their effects on the pulsation rate of Cepheids \cite{cepheids} and comparisons of x-ray and weak-lensing profiles of galaxy clusters \cite{cluster1,cluster2}. There have also been efforts to constrain the parameters of chameleon models by their effects on the cosmic microwave background \cite{CMBbeta,CMBDM}, though these analyses focus specifically on potentials of the power-law form. Given the tremendous experimental effort under way to detect or constrain chameleons, it is troubling that the most widely studied chameleon models have been shown to suffer a breakdown in calculability in the early Universe due to the discrepancy between the chameleon mass scale and that of the SM particles \cite{AdrienneShort, AdrienneLong}.

We aim to identify a chameleon potential that can avoid the computational breakdown suffered by runaway models. We analyze a class of potential not often considered in chameleon theories: the quartic potential, $V(\phi)=\kappa\phi^4/4!$. Prevalent in high-energy theories, the quartic potential is also viable as a chameleon model because the self-interaction of this potential is sufficient to ensure that the field will be adequately screened in high-density environments \cite{Phi4}. The scale-free property of the quartic model is potentially beneficial as it can avoid the hierarchy of energy scales that arises due to the low-energy scale of the runaway potentials, and we investigate whether it is able to remain well-behaved in the early Universe.\footnote{Another proposed way to avoid the detrimental effect of the kicks is to include DBI-inspired corrections to the chameleon's Lagrangian that weaken the chameleon's coupling to matter at high energies. This modification effectively introduces a second screening mechanism analogous to a Vainshtein screening in which derivative interactions weaken the effect of the kicks \cite{SwiftKick}.}

In runaway chameleon models, the field rolls to some value far from the minimum of its effective potential after inflation and remains stuck there during the radiation-dominated era due to Hubble friction. The chameleon's coupling to the trace of the stress-energy tensor makes it sensitive to the energy density, $\rho$, and pressure, $P$, of the radiation bath through the quantity $\Sigma\equiv(\rho-3P)/\rho$. While the Universe is radiation dominated, $\Sigma$ is nearly zero and the chameleon is light enough that Hubble friction is able to prevent it from rolling toward its potential minimum. However, as the temperature of the radiation bath cools, particle species in thermal equilibrium become nonrelativistic and $\Sigma$ momentarily becomes nonzero. The chameleon then gains mass, is able the overcome Hubble friction, and is seemingly ``kicked'' toward the minimum of its effective potential \cite{Brax}.

Originally, the kicks were seen as an auspicious way to bring runaway chameleons to their potential minimum prior to Big Bang Nucleosynthesis (BBN).\footnote{A consequence of the chameleon's coupling to matter is that any variation in the chameleon field can be recast as a variation in particle masses in the Jordan frame. As we know particle masses differed very little between BBN and the present day, this constrains the chameleon to be at or near the minimum of its potential prior to the onset of BBN \cite{Brax}.} However, they impart such a high velocity to the field that the chameleon rebounds off the other side of its effective potential back to field values further from the potential minimum than where it was stuck when the kick began \cite{EM}. However, Ref. \cite{EM} also showed that the inclusion of a coupling between the chameleon and the electromagnetic field offers a solution.  The chameleon's coupling to a primordial magnetic field allows the chameleon to overcome Hubble friction and begin oscillating about its potential minimum prior to the kicks. For a sufficiently rapidly oscillating field, the kicks then have little effect on the chameleon's evolution.

These kicks further jeopardized chameleon theories by throwing into question their validity as a classical field theory \cite{AdrienneShort, AdrienneLong}. The effective potential in runaway models is minimized when $\phi\sim\!M$, and at field values $\phi\lesssim\!M$, the extremely steep slope of the bare potential leads to rapid changes in the chameleon's effective mass for small field displacements. Thus, the GeV-scale velocity with which the chameleon approaches its meV-scale minimum after the kicks causes nonadiabatic changes in the mass that excite extremely energetic fluctuations and lead to the quantum production of particles \cite{AdrienneShort, AdrienneLong}. Quantum corrections due to particle production then invalidate the classical treatment of the chameleon field and the particles' trans-Planckian energies cast doubt on the chameleon's viability as an Effective Field Theory (EFT) at the energy scale of BBN.

We will show that the quartic potential is able to avoid these problems due to its scale-free nature. In the early Universe, a chameleon field in a quartic potential oscillates rapidly with a large amplitude far beyond the minimum of its effective potential. In the classical treatment, the chameleon would continue this behavior until the end of radiation domination and still be oscillating far outside its minimum at the onset of BBN. A quantum treatment of the chameleon's motion shows that these oscillations will create particles, albeit with much less energy than those created during the rebounds off runaway potentials. The same quantum effects that were catastrophic to previous chameleon models will cause the field to lose energy and bring the quartic chameleon to its potential minimum prior to the onset of the kicks. Consequently, for the quartic chameleon, these kicks do not have as significant an influence on the field's evolution as in models with runaway potentials. Depending on the value of $\kappa$, the rate at which the field loses energy can vary significantly. For large values of $\kappa$, the field can lose all of its initial energy to particle production within the first oscillation and fall to its minimum. However when $\kappa$ is closer to unity only a small percentage of the energy is lost during each oscillation, but the total effect accumulates over many oscillations to introduce a decay factor to the amplitude that still allows the field to reach its potential minimum before the kicks.

We begin with a brief review of chameleon gravity and then analyze the evolution of a classical chameleon field in a quartic potential in Section II. Then, in Section III, we consider the effects of quantum particle production on the field and investigate how the energy lost to this process is affected by the choice of $\kappa$. Section IV explores how the kicks affect the field, and we follow up with concluding remarks in Section V. Throughout this paper we will use $\Mpl=(8 \pi G)^{-1/2}$ and $c=\hbar =1$.

\section{Classical Chameleons}
\label{sec:Classical}

In theories of chameleon gravity, the action can be written as
\begin{align}
S=& \int d^4x \sqrt{-g_*} \left[\frac{\Mpl^2}{2}R_* - \frac{1}{2}(\nabla_* \phi)^2 - V(\phi)\right]  \nonumber \\
&+S_m\left[\gtil_{\mu\nu}, \psi_m\right],
\label{eq:action}
\end{align}
where $g_*$ is the determinant of the metric $g^*_{\mu\nu}$ that solves the Einstein equations, $R_*$ is its Ricci scalar, and $V(\phi)$ is the potential of the chameleon field, $\phi$. The spacetime metric $\gtil_{\mu\nu}$ that appears in the action for the matter fields, $S_m$, governs geodesic motion and is conformally coupled to the Einstein metric by
\beq
\gtil_{\mu\nu} = e^{-2\beta\phi/\Mpl}\ g^*_{\mu\nu},
\label{eq:coupling}
\eeq
where $\beta$ is a positive, dimensionless coupling constant assumed to be of order unity.\footnote{In most other chameleon theories, the bare potential and the matter coupling term must slope in opposite directions in order to produce the required minimum in the effective potential, and the coupling is generally given with a positive exponential.  However, for the quartic potential, the coupling may slope in either direction and still produce a minimum, so in following with Ref. \cite{Phi4} we will use this form of the coupling, which essentially gives a coupling constant $\beta$ that is negative compared to most theories.} This coupling implies that the Einstein-frame stress-energy tensor of the matter fields is ${T_*^\mu}_\nu = e^{-4\beta\phi/\Mpl}\tilde{T}^\mu{}_\nu$. With this relationship between ${T_*^\mu}_\nu$ and $\tilde{T}^\mu{}_\nu$, the Einstein and Jordan frame energy density and pressure can be related by $\rho_*/\tilde{\rho}=P_*/\tilde{P}= e^{-4\beta\phi/\Mpl}$. It follows that any quantity that is a ratio of elements of the stress energy tensor, such as $\Sigma$ or $w\equiv P/\rho$, is the same in both frames and can be evaluated using either Einstein- or Jordan- frame quantities.

Varying the action with respect to $g^*_{\mu\nu}$ implies that ${T_*^\mu}_\nu$ is not conserved in the Einstein frame, as energy is exchanged between matter and the chameleon field. However, as the scalar and matter fields do not interact in the Jordan frame, the Jordan-frame stress-energy tensor is conserved: $\tilde{\nabla}_\mu \tilde{T}^\mu{}_\nu = 0$. In a Friedmann-Robertson-Walker spacetime, the scale factors in the Jordan and Einstein frames are related by $\tilde{a}=e^{-\beta\phi/\Mpl}a_*$ and the proper times are related by $d\tilde{t}=e^{-\beta\phi/\Mpl}dt_*$. Since the Einstein-frame matter density is not conserved, it does not follow the usual $a_*^{-3}$ scaling. Radiation, however, still follows the expected $a_*^{-4}$ behavior, as we show in the Appendix. Throughout the remainder of the paper we will drop the $*$ subscript on the scale factor when discussing how quantities scale in the Einstein frame.

The relationship between ${T_*^\mu}_\nu$ and $\tilde{T}^\mu{}_\nu$ also implies that the Jordan-frame temperature, $T_J$, depends on $\phi$. As entropy is conserved in the Jordan frame, $g_{*S}(T_J) \tilde{a}^3 T_J^{3}$ is constant, and the expression for $T_J$ in terms of $\phi$ and $a_*$ is
\beq
T_J \left[{g_{*S}(T_J)}\right]^{1/3}= \left[{g_{*S}(\TJi)}\right]^{1/3} \TJi e^{-\beta(\phi_i-\phi)/\Mpl} \frac{a_{*,i}}{a_*}.
\label{T_J}
\eeq
where $a_{*,i}$ is the initial value of $a_*$, $\phi_i = \phi(a_{*,i})$, and $\TJi = T_J(a_{*,i})$.

\subsection{Chameleon Cosmology}
\label{sec:Cosmology}

Varying the action with respect to the field $\phi$ gives the equation of motion for the chameleon:
\begin{align}
\ddot{\phi}+3H_* \dot{\phi} =& - \frac{dV}{d\phi}-\frac{\beta}{\Mpl}{T_*^\mu}_\mu ;\\
=& - \frac{dV}{d\phi} + \frac{\beta}{\Mpl}(\rho_*-3P_*),
\label{eq:eom}
\end{align}
where the dot denotes a derivative with respect to proper time $t_*$ in the Einstein frame and $H_* \equiv\dot{a}_*/a_*$.

The effective potential that controls the evolution of the chameleon field is
\begin{align}
\Veff(\phi) & = V(\phi)-\frac{\beta\phi}{\Mpl} (\rho_*-3P_*);
\label{eq:VgenEff}\\
 & = \frac{\kappa}{4!}\phi^4-\frac{\beta\phi}{\Mpl} \Sigma\rho_{*},
\label{eq:Veff}
\end{align}
where $\kappa$ is a dimensionless constant, and we have used the definition $\Sigma\equiv(\rho_*-3P_*)/\rho_*$. Quantum loop corrections to the classical potential and limits on fifth forces constrain $\beta$ and $\kappa$. The chameleon mechanism depends on an increase in the chameleon's effective mass in order to hide its effects, however quantum corrections to its potential also increase with its mass. Maintaining the reliability of fifth-force predictions requires that these corrections remain small compared to the classical potential and places an upper limit on the chameleon mass that implies $\kappa \lesssim 100$ \cite{Quantum}. Laboratory searches for fifth forces, in turn, have already placed lower bounds on the chameleon mass, which can to used to constrain $\kappa$ from below for given $\beta$ \cite{EotWash15}. In order for $\kappa$ to be of order unity, the chameleon coupling must be $\beta \lesssim 10^{-1}$. Conversely, in order for $\beta$ to be of order unity, $\kappa$ must be $\gtrsim 50$.

The minimum of this effective potential,
\beq
\phimin=\left(\frac{6\beta \Sigma\rho_*}{\kappa\Mpl}\right)^{1/3},
\label{eq:min}
\eeq
is dependent on $\rho_*$ and on $P_*$ through the definition of $\Sigma$, and so, too, is the chameleon's effective mass \mbox{$m^2=\left.d^2V/d\phi^2\right|_{\phi=\phi_{\rm min}}$}. The mass increases with $\rho_*$, making the chameleon heavier in regions of high density and unable to mediate a long-range force.

When evaluating the chameleon equation of motion, we work with a dimensionless scalar field $\varphi\equiv\phi/\Mpl$, as well as with $p\equiv \mathrm{ln}(a_*/a_{*,i})$. Primes will now denote differentiation with respect to this new time variable $p$ and the first Friedmann equation is
\beq
H_*^2=\frac{\rho_*+V}{3\Mpl^2\left[1-(\varphi^\prime)^2/6\right]}.
\eeq
Using the above equation and the fact that $\Sigma\ll1$ during radiation domination and \mbox{$V(\phi)\ll\rho_*$}, the chameleon equation of motion, Eq.~(\ref{eq:eom}), can be written as
\beq
\frac{\rho_*+V}{\left[1-(\varphi^\prime)^2/6\right]}\varphi^{\prime\prime}= -\varphi^\prime\left(\rho_*+3V\right)-3\left(\frac{dV}{d\varphi}-\beta\Sigma\rho_*\right).
\label{eq:fulleom}
\eeq
We will use this equation in order to explore the evolution of a chameleon field in a quartic potential throughout the radiation-dominated era.

The initial conditions for Eq.~(\ref{eq:fulleom}) follow from the field's dynamics prior to reheating. During inflation, the equation of state parameter $w$ is approximately $-1$, and the comparatively large value of the kick function, $\Sigma=(1-3w)\simeq4$, sets the value of $\phimin$ drastically greater than it is during radiation domination. The mass of the field at its minimum is
\beq
m^2 = \frac{\kappa}{2}\phimin^2 = \left(\frac{9}{2}\kappa\beta^2\right)^{1/3}\left(\frac{\Sigma\rho_*}{\Mpl}\right)^{2/3}.
\eeq
When $\Sigma\gtrsim1$, the response time of the field $m^{-1}$ is much shorter than the Hubble time $H_*^{-1}$ as long as $\rho_* \ll \Mpl^4$,
\begin{align}
\frac{m^2}{H_*^2} & \simeq 3 \left(\frac{9}{2}\kappa\beta^2\right)^{1/3} \left(\frac{\Sigma^2\Mpl^4}{\rho_{*}}\right)^{1/3}.
\label{eq:m2h2infl}
\end{align}
Therefore, the field is massive enough to roll to its minimum prior to the onset of radiation domination. The fact that $m^2 \gg H^2$ also implies that the chameleon field is massive enough during inflation that quantum effects do not generate superhorizon perturbations in its value.

During reheating, the energy density $\rho_*$ (be it of the inflaton or another oscillating scalar field) is converted into radiation. The value of the kick function then drops to $\Sigma\ll1$, and $\phimin$ is pushed to significantly smaller field values; see Eq.~(\ref{eq:min}). For all reheat temperatures much less than $\Mpl$ we can assume that the chameleon begins at rest with $\phi$ equal to the value of $\phimin$ just prior to the drop in $\Sigma$, because $m^2 \gg H^2$, as shown in Eq.~(\ref{eq:m2h2infl}). At temperatures greater than a TeV, the QCD trace anomaly implies that the value of $\Sigma$ is 0.001 \cite{Curvature}. $\Sigma$ then maintains this value until $T_J\simeq600$ GeV when contributions from massive particles become comparable. As the temperature decreases, $\Sigma$ begins to increase as $\Sigma\propto m_t^2/T^2$, where $m_t$ is the mass of the most massive SM particle species: the top quark \cite{Curvature}.

At $T_J\simeq200$ GeV the process that gives the kick function its name begins. As the temperature of the radiation bath decreases, the energy density and pressure of massive particles decay at slightly different rates, allowing $\Sigma$ to reach non-negligible values. This happens as each SM particle becomes nonrelativistic, but contributions from some species merge together, and the entire process results in four distinct kicks. The contributions from each particle are suppressed by a factor of $g_*(T_J)^{-1}$, where $g_*(T_J)$ is the effective degrees of freedom. As the temperature cools, $g_*(T_J)$ decreases, and each kick becomes larger than the last with the final kick due to the electrons reaching a value of $\Sigma\simeq0.1$. For a detailed calculation of the kick function $\Sigma$, see Appendix A of Ref. \cite{AdrienneLong}.

\subsection{Quartic Chameleons}
\label{sec:quartic}

For the runaway potentials usually considered in chameleon gravity, the value of $V(\phi)$ approaches infinity as $\phi \rightarrow 0$ and drops off rapidly as $\phi$ increases. The effective potential is then dominated by $V(\phi)$ near $\phi=0$ and by the linear matter-coupling term at field values greater than $\phimin$. During inflation, the large value of $\Sigma$ makes the slope of the matter-coupling term in Eq.~(\ref{eq:VgenEff}) steeper, and the chameleon sits in a potential minimum at a small $\phi$ value. When inflation ends and $\Sigma$ decreases, the slope of the matter contribution to $\Veff$ becomes shallow and the minimum of the effective potential moves to larger values of $\phi$. The chameleon then rolls down its bare potential, past the minimum, and out to where the effective potential is dominated by the matter-coupling term. The field then becomes stuck due to Hubble friction until it is kicked back toward the minimum of its effective potential.

The quartic chameleon, however, feels the effects of its bare potential on both sides of the minimum of its effective potential. As previously discussed, the comparatively large value of $\Sigma$ prior to reheating fixes $\phimin$ at a large value far from zero. Throughout this analysis, we use the subscript $i$ to indicate the value of a quantity just prior to the onset of radiation domination, which we take to occur at a Jordan-frame temperature $T_{J,i}=10^{16}$ GeV. The value of $\phimin$ is then
\beq
\phi_i \simeq 0.0062 \Mpl \left(\frac{\beta}{\kappa}\right)^{1/3}.
\label{eq:phii}
\eeq
We have also assumed an era of inflation prior to radiation domination and have therefore used $\Sigma_i=4$. However, nonstandard histories can easily be accounted for by changing this value in Eq.~(\ref{eq:min}). We will show that the values of $\Sigma_i$ and $T_{J,i}$, only set the initial oscillation amplitude of the field, to which the subsequent evolution is largely insensitive.

When the value of $\Sigma$ drops from $4$ to $0.001$ at the end of inflation, the value of $\phimin$ decreases by a factor of $(0.001/4)^{1/3}\simeq 0.06$. The chameleon then rolls rapidly down its bare potential toward this new minimum. In its decent to its potential minimum, the chameleon gains sufficient energy that the slight tilt at the bottom of the quartic well due to the now small matter coupling does very little to affect its motion as it passes through $\phimin$ and climbs up the other side of its bare potential. It climbs to almost the same potential value as it started before turning around and falling again with nearly the same energy. It continues in this fashion, oscillating back and forth, all but oblivious to the matter coupling.

The oscillation amplitude decreases as Hubble friction causes the energy in the chameleon field to redshift away as $a^{-4}$. This behavior can be understood easily by the virial theorem. For a general power-law potential of the form $V(\phi)=C\phi^{n}$, the virial theorem relates the rapidly oscillating field's average kinetic and potential energies, $K$ and $V$, by
\beq
2\Kbar = n \Vbar.
\eeq
The equation-of-state parameter $w$ is then
\beq
w = \frac{\Pbar}{\rhobar} = \frac{\frac{1}{2}\phidot^2 - V}{\frac{1}{2}\phidot^2 + V} = \frac{n-2}{n+2}.
\eeq
Using this value of $w$, we can determine how the chameleon energy will scale with expansion by using the conservation equation:
\begin{align}
\rho & = \rho_0 a^{-3(1+w)};\nonumber \\
& = \rho_0 a^{-3(\frac{2n}{n+2})}.
\label{eq:virialenergy}
\end{align}
For a quartic potential, $n=4$ and the last line implies that the energy scales as $a^{-4}$. Technically, the chameleon's energy does not exactly obey Eq.~(\ref{eq:virialenergy}) because there is a small amount of energy exchanged between the field and matter. However, we show in the Appendix that the corrections to the evolution of the chameleon's energy density are negligible. We also show in the Appendix that the energy density in radiation will scale in the same way as the chameleon energy $\rho_{*R}\propto a^{-4}$, and so $\rho_\phi/\rho_{*R}\sim V_i/T_{J,i}^4\ll1$.

\begin{figure}
\centering\includegraphics[width=3.4in]{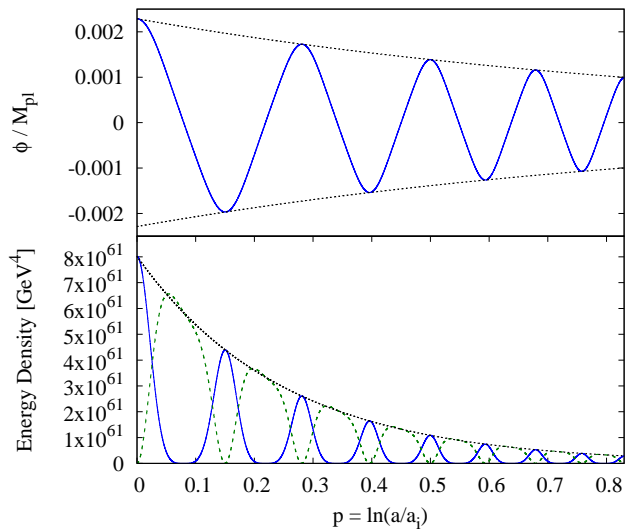}
\caption{Top: The value of $\varphi$ (blue, solid) and the values of its minima and maxima, $-\varphi_i a^{-1}$ and $\varphi_i a^{-1}$, respectively (black, dotted) over the course of four oscillations. Bottom: The kinetic (green, dashed) and potential (blue, solid) energies of the chameleon and their amplitude, $V_i a^{-4}$ (black, dotted). The potential energy reaches a maximum twice during each complete oscillation when $|\varphi|$ is at a maximum. The kinetic energy reaches its maxima both times $\varphi$ passes through $\varphimin$. At all times the chameleon energy density is much less than that of the radiation $\rho_{*R}\sim (T_{J,i}a^{-1})^4 = 10^{64}a^{-4}$GeV (In this and all figures $\beta= 0.1$ and $\kappa=2$.)}
\label{KandV}
\end{figure}

As the energy in the chameleon field is the sum of its kinetic and potential energies, the maximum values of both of these quantities during each oscillation will scale as $a^{-4}$. The potential energy of the field when it reaches the peak of each oscillation, $\Vmax$, and its kinetic energy each time the field passes through the minimum of its potential, $\Kmax = \phidotmax^2/2$, are both related to the field's initial potential energy by $\Vmax = \Kmax = V_i a^{-4}$. The quartic relation between $\phi$ and $V$ implies that the amplitude of the $\phi$ oscillations decays as $a^{-1}$, so the value of $\phi$ at the peak of each oscillation is $\phimax = \phi_i a^{-1}$. Both of these behaviors can be seen in Figure 1, which shows the value of $\varphi$ in the top panel and the kinetic and potential energy of the field in the bottom panel plotted over the course of several oscillations. These plots are generated from the numerical solution to Eq. (\ref{eq:fulleom}) assuming $\phiprime_i=0$ and $\varphi_i= \phi_i/\Mpl$ with $\beta=0.1$ and $\kappa=2$.

The fact that the quartic chameleon begins at \mbox{$\phi_i \ll \Mpl$} and does not exceed this value is an interesting difference compared to runaway models. The field value at which runaway models become stuck due to friction can be nearly equal to $\Mpl$ \cite{Brax}. If the field remains stuck at such values until BBN, the large variation of $\phi$ from its potential minimum can be interpreted as a larger variation in particle masses than we know to be allowed. Quartic chameleons, however, are already at field values much less than $\Mpl$ before the end of inflation and oscillate with a decreasing amplitude. While we will show that the field still finds its minimum prior to the kicks, it is not strictly necessary to avoid endangering the success of BBN.

Equation (\ref{eq:min}) implies that $\phimin$ is proportional to the cube root of the energy density in radiation and so will decay as $a^{-4/3}$. Thus, $\phimin$ will decrease faster than the oscillation amplitude by a factor of $a^{-1/3}$, implying that the value of $\phi$ at the maximum of its oscillations will always exceed the minimum of its effective potential.  Therefore, our classical treatment of the chameleon's behavior suggests that it would spend most of its time in regions dominated by its bare potential far from the minimum of its effective potential (though not far enough to significantly effect particle masses), allowing the oscillations to continue indefinitely while the Universe is radiation dominated. The high-energy oscillations of the field prevent it from becoming stuck due to Hubble friction or falling into and tracking its minimum. However, as the problems with other chameleon models demonstrate, the quantum effects associated with rapid changes in the chameleon field can significantly alter this classical behavior.

\section{Quantum Chameleons}
\label{sec:quantum}

In chameleon models with runaway potentials, the only instances of rapid changes of the chameleon field after inflation occur when the chameleon is kicked toward its potential minimum with a very high velocity and rebounds off its steep bare potential. The rapid changes in the mass of the chameleon during this rebound excite high-energy perturbations that, in a naive, classical evaluation, exceed the energy initially available to the chameleon field. Considerations of the backreaction of particle production on the field showed that quantum corrections significantly alter the form of the potential experienced by the chameleon field. These corrections radically change the chameleon's evolution throughout the rebound, causing it to turn around long before it would have exhausted the kinetic energy it possessed going into the rebound, which keeps the occupation numbers of the excited modes extremely small \cite{AdrienneLong}.

In this section we show that every oscillation of the quartic chameleon excites perturbations, but with small enough energies that the energy lost to particle production does not exceed the initial energy of the field. For increasing values of $\kappa$ we find that the limit at which this is no longer the case coincides with the results Ref. \cite{Quantum}, which also used quantum corrections to place an upper bound on $\kappa$. For the relatively large values of $\kappa$ near this limit, the field can lose all of its initial energy to particles before it completes an oscillation, and it simply falls to its  potential minimum. For smaller values ($\kappa \lesssim {\cal O}(1)$), the energy lost is only a small fraction of the field's energy at the start of an oscillation, and the evolution of the field over a single oscillation is not significantly altered. Instead, this energy loss accumulates over many oscillations and introduces an additional decay factor to the oscillation amplitude causing it to decay faster and reach its potential minimum.

\subsection{Particle Production}
\label{sec:particles}

We first summarize how rapid changes in the chameleon's effective mass excite perturbations \cite{Birrell}; for a more detailed review of this process, see Appendix C of Ref. \cite{AdrienneLong}. We begin by decomposing the field into its spatial average $\phibar(t)$ and the perturbation $\delta \phi$:
\beq
\phi(t,{\bf x}) = \phibar(t) + \delta\phi(t,{\bf x}).
\label{eq:split}
\eeq
The linearized perturbation equation that governs the evolution of $\delta\phi$ is
\beq
\left[ \partial_t^2 + 3 H \partial_t - \frac{\nabla^2}{a^2} + V_{\mathrm{eff}}^{\prime\prime}(\phibar) \right] \delta \phi = 0.
\label{eq:perteq}
\eeq

Throughout this section we will not be using the variable $p$, and primes will denote differentiation with respect to the argument of the function.

To quantize the perturbations, we introduce the creation and annihilation operators $\hat{a}^\dagger_{\bf k}$ and $ \hat{a}_{\bf k}$, respectively, which obey the standard commutation relations,
\beq
\left[\hat{a}_{\bf k} , \hat{a}_{\bf k'}^\dagger \right] = (2\pi)^3\delta^{(3)}\left({\bf k}-{\bf k'}\right).
\eeq
The annihilation operator annihilates the vacuum state: $\hat{a}_{\bf k} |0\rangle = 0$. Using $\hat{a}^\dagger_{\bf k}$ and $ \hat{a}_{\bf k}$ we can then express $\delta\phi(\tau)$ as
\beq
\hat{\delta\phi}(\tau,{\bf x}) = \int \frac{d^3k}{(2\pi)^{3}} \left[ \hat{a}_{\bf k} \frac{\phi_k(\tau)}{a(\tau)} e^{i{\bf k}\cdot {\bf x}} +\hat{a}^\dagger_{\bf k} \frac{\phi^*_k(\tau)}{a(\tau)} e^{-i{\bf k}\cdot {\bf x}}  \right],
\label{eq:delphiquant}
\eeq
where $\tau$ is conformal time. Inserting this decomposition of $\delta\phi$ into Eq. (\ref{eq:perteq}), we find
\begin{align}
& \phi_k''(\tau) + \omega_k^2(\tau) \phi_k = 0; \\
& \omega_k^2(\tau) = k^2 + a^2 V_{\mathrm{eff}}''(\phibar) - \frac{a''(\tau)}{a},
\end{align}
where $\omega_k^2(\tau)$ is the effective mass of a plane-wave perturbation in the chameleon field with a comoving wavenumber $k$.

During radiation domination $a''(\tau)=0$, and
\beq
\omega_k^2 = k^2 + a_*^2 V_{\mathrm{eff}}''\left(\phibar\right)\simeq k^2 + \frac{\kappa}{2}(a_*\phi)^2,
\label{eq:omegasq}
\eeq
where in the last equality we have dropped the bar over $\phi$ as we will be working exclusively with the spatially averaged field. We also neglect the matter coupling because it is subdominant to the bare potential throughout most of the oscillation.  When $\omk'(\tau) / \omk^2 \gtrsim 1$, perturbations in the field are excited. Taking the derivative of this effective mass with respect to $\tau$, we find
\begin{align}
\omk'(\tau)  & = \frac{a_*^3}{2\omk}\left[2 H_* V^{\prime\prime}(\phi) + V^{\prime\prime\prime}(\phi)\dot{\phi}\right];\nonumber\\
& = \frac{a_*^3}{2\omega_k}\left[\kappa H_* \phi^2 + \kappa\phi\phidot\right].
\label{eq:omegaprime}
\end{align}

We can simplify the last line of the equation by noting that not only is $H_* \phi^2$ initially smaller than $\phi\phidot$, it also redshifts away faster. This can be seen by recalling the relations $\phimax = \phi_i a^{-1}$, $\phidotmax = \sqrt{2 \Kmax} = \sqrt{2 V_i a^{-4}}$, and using the fact that, during radiation domination, $H_*$ decreases as $a^{-2}$. With these results, the maximum value of the first term during each oscillation is $H_* \phimax^2 \simeq H_{*,i} \phi_i^2 a^{-4}$. The second term, however, is a product of two oscillating functions that reach their maxima at different times. From the approximately sinusoidal nature of $\phi$, we can determine that the product of $\phi$ and its derivative $\phidot$ will behave as the product of their amplitudes and another sinusoidal function, thus, $(\phi\phidot)_\mathrm{max}$ is proportional to $\phimax\phidotmax$. The constant of proportionality can be numerically determined and is $\approx\!\!0.6$ for all $\beta$ and $\kappa$. The amplitude of the second term is then $(\phi\phidot)_\mathrm{max} \sim \phimax \phidotmax \simeq \phi_i a^{-1} \sqrt{2 \Kmax} \simeq \sqrt{\kappa/12}\phi_i^3 a^{-3}$. We can see that $H_* \phi^2 \propto a^{-4}$ will decay away faster than $\phi\phidot \propto a^{-3}$. Thus, if $H_* \phi^2$ is initially the smaller of the two terms, we can neglect it.

During radiation domination, $H_*^2 \simeq \rho_{*R}/(3\Mpl^2)$, and from Eq.~(\ref{eq:min}), we know that $\phi_i = [24\beta\rho_{*R,i}/(\kappa\Mpl)]^{1/3}$. Assuming that $T_i = 10^{16}$ GeV, the relative contribution between the two terms is $\sqrt{12/\kappa}H_{*,i}/(0.6\phi_i) \simeq 0.05 (\kappa\beta^2)^{-1/6}$. Thus, as long as $\kappa\beta^2 > 1.56\times10^{-8}$, which is true provided that neither $\beta$ nor $\kappa$ is unreasonably small, this ratio is less than 1 and we can neglect the $H_*\phi^2$ term in Eq.~(\ref{eq:omegaprime}).

By setting the ratio $\omk'(\tau)  / \omk^2$ equal to $1$, we can find the physical wavenumbers, $\kphys = k/a_*$, of the perturbations that are excited during the oscillations.
Using Eqs.~(\ref{eq:omegasq}) and (\ref{eq:omegaprime}), we now have

\begin{align}
\frac{\omk'(\tau) }{ \omega_k^2(\tau) } & \simeq \frac{ a_*^3}{2\omk^3}\left(\kappa \phi\phidot\right), \nonumber\\
& = \frac{a_*^3}{2}\frac{\kappa\phidot\phi}{\left[k^2+\frac{\kappa}{2}(a_*\phi)^2\right]^{3/2}}, \nonumber\\
& = \frac{\kappa\phidot\phi}{2\left[\kphys^2+\frac{\kappa}{2}\phi^2\right]^{3/2}}.
\end{align}
Setting this ratio equal to $1$, and solving the last line for $\kphys$, we get
\beq
\kphys^2 = \left(\frac{\kappa}{2}\phi\phidot\right)^{2/3} - \frac{\kappa}{2}\phi^2.
\label{eq:kphys}
\eeq
This expression reaches its maximum value four times during a single oscillation, as shown in Figure 2.

To evaluate $\kphys$ we again use how the maxima of the contributing quantities are related to the initial field value and corresponding initial potential energy. Already we have established that, for the first term, $(\phi\phidot)_\mathrm{max} = A\phimax \phidotmax$, where $A \simeq 0.6$. It is also important to note that the terms in Eq.~(\ref{eq:kphys}) do not reach their maxima at the same time $\kphys$ is at its maxima. To account for these proportionalities we introduce the numerical parameters $B$ and $C$ to relate the maxima of the two terms to their values when $\kphys$ is at its maxima. The maximum value of $\kphys$ during an oscillation is then
\begin{align}
\left(\kphys^{\mathrm{max}}\right)^2 & = {B}\left(\frac{\kappa}{2}A\phimax\phidotmax\right)^{2/3} - C\frac{\kappa}{2}\phimax^2 \nonumber\\
& = A^{2/3}B\left(\frac{\kappa}{2}\phi_i a^{-1}\sqrt{\frac{\kappa}{12}\phi_i^4 a^{-4}}\right)^{2/3} - C\frac{\kappa}{2}\phi_i^2a^{-2} \nonumber\\
& = \frac{\kappa}{2}\left[\left(\frac{A^2}{6}\right)^{1/3}B-C\right]\phi_i^2a^{-2} \nonumber\\
\kphys^{\mathrm{max}} & = D \sqrt{\frac{\kappa}{2}}\phi_i a^{-1}
\label{eq:kphysphi}\\
& = D (6\kappa)^{1/4}V_i^{1/4} a^{-1}.
\label{eq:kphysV}
\end{align}
For $\beta=0.1$ and $\kappa=2$, $B\simeq 0.6$, $C\simeq0.1$, and as it is defined in the last equation, $D \simeq 0.4$. Changing $\beta$ or $\kappa$ by two orders of magnitude does not significantly effect the value of $A$, $B$, $C$, or $D$.

\begin{figure}
\centering\includegraphics[width=3.4in]{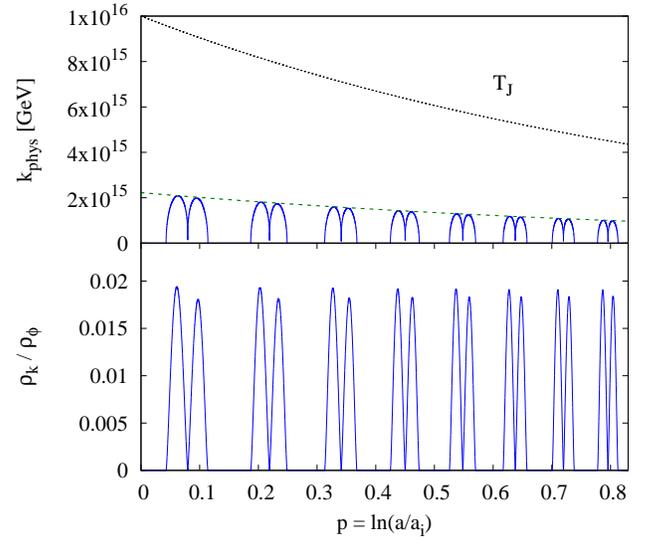}
\caption{Top: The value of $\kphys$ (blue, solid) over four oscillations. During each oscillation, $\kphys$ reaches a maximum four separate times at approximately the middle of the climb and decent on each side of the potential. The value of this maximum, $\kphys^{\mathrm{max}}$ (green, dashed), is on the order of the temperature  and decays as $a^{-1}$. Bottom: The ratio of the energy in the perturbations to the energy in the chameleon field over four oscillations.  Each oscillation sees approximately a total 0.08 fractional loss of energy.}
\label{kphys}
\end{figure}

As well as being proportional to the oscillation amplitude, the $a^{-1}$ behavior of $\kphys^{\mathrm{max}}$ implies that it also scales with the temperature. Starting from Eq.~(\ref{eq:kphysphi}), the relationship between $\kphys^{\mathrm{max}}$ and the Jordan-frame temperature becomes apparent when $\phi_i$ is determined by evaluating Eq.~(\ref{eq:min}) using $\Sigma = 4$ and $\rho_{*,i} = (\pi^2/30)g_*(T_{J,i})T_{J,i}^4$. After combining all numerical factors, we find that
\begin{align}
\kphys^{\mathrm{max}} & \simeq 2.67 (\kappa\beta^2)^{1/6} \left(\frac{(T_{J,i})^4}{\Mpl}\right)^{1/3}a^{-1}, \nonumber\\
& \simeq 2.67 (\kappa\beta^2)^{1/6} \left(\frac{T_{J,i}}{\Mpl}\right)^{1/3} T_J,
\end{align}
where we have used the fact $T_J \propto a^{-1}$ during radiation domination.\footnote{Though the temperature does depend on $\phi$, variations in the field have a negligible effect on the temperature as long as $\phi \ll \Mpl$, as the lack of deviation from the expected $a^{-1}$ behavior in the temperature plotted in Figure \ref{kphys} shows.} For $T_{J,i} = 10^{16}$ GeV, $\beta=0.1$, and $\kappa=2$, the ratio of $\kphys$ to the temperature is $\kphys^{\mathrm{max}} / T_J \simeq 0.23$, as shown in Figure 2.

The fact that the energy of the modes excited in quartic models is dependent only on the initial value of the field is another important contrast to runaway models. The energy of excited modes in such models is dependent on the velocity with which the chameleon approaches the minimum, $\vel$, which is of order GeV$^2$. Reference \cite{AdrienneLong} found that for a power-law potential of the form
\beq
V(\phi) = M_v^4\left[1+\left(\frac{M_s}{\phi}\right)^n\right],
\label{Vpow}
\eeq
the most energetic mode that is excited has a physical wave number
\beq
\kphys =  \frac{(n+2)}{2\sqrt{2}} \frac{|\vel|}{M_s}\left(\frac{M_s}{\phita}\right).
\label{kex}
\eeq
where $\phita \lesssim M_s$ is the value of $\phi$ at which the field turns around. Since $M_s \sim$ meV, the ratio of $\vel \sim T_J^2$ and $M_s$ results in the excitation of extremely energetic modes even at low temperatures: $\kphys \gg \sqrt{\vel} \sim T_J$. For the quartic chameleon, however, highly energetic modes are only excited at high temperatures: $\kphys \lesssim T_J$.

Using our value of $\kphys$, we can evaluate the energy density in the perturbations:
\begin{align}
\rho_k = \frac{k^3n_k\omk}{2\pi^2 a^4} & \simeq \frac{\kphys^3}{2\pi^2}\sqrt{\kphys^2 + V^{\prime\prime}(\phi)}, \nonumber\\
& = \frac{\kphys^3}{2\pi^2}\sqrt{\left(\frac{\kappa}{2}\phidot{\phi}\right)^{2/3} - \frac{\kappa}{2}\phi^2 + \frac{\kappa}{2}\phi^2}, \nonumber\\
& = \frac{\kphys^3}{2\pi^2}\left(\frac{\kappa}{2}|\phi\phidot|\right)^{1/3},
\end{align}
where $n_k \sim 1$ is the mode occupation number. In the same way that we found $\kphys^{\mathrm{max}}$ by relating the maxima of the quantities in Eq.~(\ref{eq:kphys}) to the initial potential energy, we can find the maximum of $\rho_k$ during each oscillation,
\beq
\rho_k^{\mathrm{max}} = \frac{D^3}{2\pi^2}6^{5/6}\kappa V_i a^{-4}.
\label{eq:rhokmax}
\eeq
From Section II we know that the chameleon's energy density equals $V_i a^{-4}$. Therefore, the maximum of the ratio of $\rho_k$ and the energy at the start of an oscillation $\rho_\phi$ is constant, as we can see in the bottom panel of Figure \ref{kphys}. The ratio of these two quantities is dependent on $\kappa$: $\rho_k/\rho_\phi \simeq 0.01\kappa$. To ensure our quantum corrections are kept under control, this ratio must be $<1$, and we must have $\kappa\lesssim100$. This is the same bound found by Ref. \cite{Quantum}, which used another approach to limit quantum corrections. Runaway models could only keep this ratio less than 1 if the occupation number was extremely small, $n_k \lll 1$, which required altering the classical evolution of the field.

For $\kappa \gtrsim {\cal O}(10)$ the chameleon field will lose all of its energy before it has a chance to complete its first oscillation, at which point it will settle into its potential minimum. For smaller values of $\kappa$, however, the depreciation in the field's energy is not as dramatic, and the small fractional loss of energy does not significantly affect the field's evolution during a single oscillation. Instead, the effect accumulates over many oscillations, as we explore in the next section.

\subsection{Effects of Particle Production for $\kappa \lesssim {\cal O}(1)$}
\label{sec:effectparticles}

While the fraction of the energy lost in each oscillation is constant, the length of each oscillation period, $\delp$, is not, as can be seen in Figure 2. The duration of the oscillations scales as follows:
\begin{align}
\delp & \simeq \frac{4 \varphimax(p)}{\phiprimeavg},\nonumber\\
& \simeq \frac{4 \varphi_i e^{-p}}{\phiprimeavg},
\label{eq:delp}
\end{align}
where $\phiprimeavg$ is the average value of $\phiprime$ over a single oscillation and once again primes denote differentiation with respect to $p$. To more clearly see the behavior of $\delp$, we first remark on the quantity $\varphi^\prime = \phidot/(\Mpl H_*)$. Radiation domination implies that $H_*$ will decrease as $a^{-2}$, and the fact that the maximum kinetic energy of the chameleon during an oscillation, $\phidotmax^2/2$, is proportional to $a^{-4}$ implies that $\phidotmax$ also decreases as $a^{-2}$.  Therefore, the amplitude of the $\varphi^\prime$ oscillations is a constant value: $\phiprimemax$. Since $\phiprimemax$ is constant, so too is its average, and $\delp$ decays with the amplitude of $\varphi$. The two values $\phiprimeavg$ and $\phiprimemax$ can be related by a constant scaling factor found numerically to be $q\equiv\phiprimeavg/\phiprime_\mathrm{max}\simeq 0.76$, and is highly insensitive to changes in $\beta$ and $\kappa$ of up to 2 orders of magnitude.

When considering the energy lost during each oscillation, it is useful to consider how the quantities $\varphimax$ and $\phiprimeavg$ are related to the energy at the start of each oscillation, $\rho(p)$:
\begin{align}
\varphimax(p) & = \left(\frac{4!}{\kappa}\frac{\rho(p)}{\Mpl^4}\right)^{1/4}
\label{eq:phimax}\\
\phiprime_\mathrm{max}(p) & = \frac{\phidot_\mathrm{max}}{H_*\Mpl} = \frac{\sqrt{2 \rho(p)}}{H_*\Mpl} = \sqrt{\frac{\kappa}{12}}\frac{\Mpl}{H_*}\varphimax^2(p),
\label{eq:phiprimemax}
\end{align}
where we have used the fact that the maximum kinetic energy that occurs during each oscillation, $\phidot_\mathrm{max}^2/2$, is equal to the potential energy at the start of each oscillation. Using Eqs.~(\ref{eq:phimax}) and ~(\ref{eq:phiprimemax}) and the scaling constant $q$, Eq.~(\ref{eq:delp}) then becomes
\begin{align}
\delp & = \frac{4}{q}\sqrt{\frac{12}{\kappa}}\frac{H_*}{\Mpl}\varphimax^{-1},\nonumber\\
& = \frac{4}{q}\left(\frac{6}{\kappa}\right)^{1/4}H_i e^{-2p} \rho^{-1/4},\\
& = \frac{1}{Q} e^{-2p}\rho^{-1/4},
\label{eq:period}
\end{align}
where $H_i$ is the initial Hubble value at the end of inflation and in the last line we have condensed all the constants into one constant, $Q^{-1}$.

\begin{figure}
\centering\includegraphics[width=3.4in]{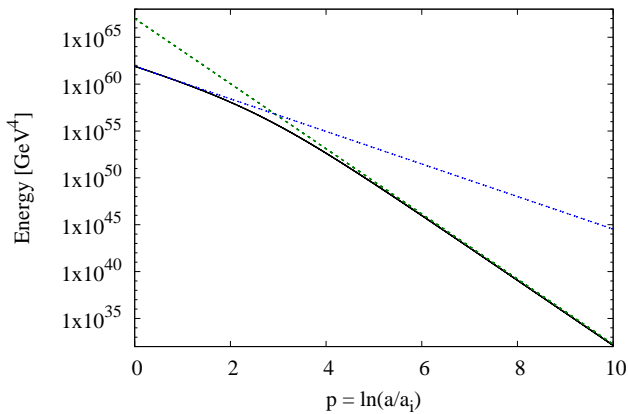}
\caption{The energy in the chameleon field at the peak of each oscillation (black, solid), showing the effect of particle production on the energy decay, found by numerically solving Eq.~(\ref{eq:energydecay}). While the energy begins redshifting as $a^{-4}$ (blue, dotted), particle production eventually dominates the energy loss and the energy decays as $a^{-8}$ (green, dashed).}
\label{energydecay}
\end{figure}

Over the course of each oscillation, the energy of the chameleon field decreases by a factor of $e^{-4\delp}$ due the expansion of the Universe, as well as by an additional factor due to the creation of particles. If we take $f \lesssim 1$ as the fraction of energy left after the field has lost energy due to the production of particles during one oscillation, we can write the change in energy over one oscillation period as
\begin{align}
\frac{\Delta\rho}{\delp} & = \frac{f\rho_\mathrm{init}e^{-4\delp}-\rho_\mathrm{init}}{\delp},\nonumber\\
& \simeq \frac{f\rho_\mathrm{init}(1-4\delp)-\rho_\mathrm{init}}{\delp}.
\label{eq:energydelta}
\end{align}
If $f=1$ we recover the original $\rho \propto e^{-4p}$ evolution. Using Eqs.~(\ref{eq:period}) and (\ref{eq:energydelta}) we can write a differential equation for the energy loss including particle production,
\begin{align}
\frac{d\rho}{dp} & = -\rho\left[\frac{1-f}{\delp} + 4f\right] \nonumber\\
& = - Q(1-f) e^{2p}\rho^{5/4} - 4f\rho
\label{eq:energydecay}
\end{align}
The second term in Eq.~(\ref{eq:energydecay}) gives the classical $\rho \propto a^{-4}$ evolution and dominates at small $p$. But the $e^{2p}$ factor in the first term allows it to quickly dominate over the second term. If we consider the regime in which the second term has become negligible, we can integrate Eq.~(\ref{eq:energydecay}) and see that the energy will follow an entirely different behavior at late times:
\begin{align}
\frac{d\rho}{dp} & = - Q(1-f) e^{2p}\rho^{5/4}\nonumber\\
\int \frac{d\rho}{\rho^{5/4}} & = -Q(1-f)\int e^{2p}dp \nonumber\\
\rho & = \left[\frac{Q(1-f)}{8}e^{2p} + C\right]^{-4}
\end{align}
where C is a constant of integration. At large values of $p$, when this behavior is relevant, the exponential term will dominate over the constant and the energy will scale as $e^{-8p}$. The two regimes and behavior of $\rho$ can be seen in Figure \ref{energydecay}. The $p$ value at which the $e^{-8p}$ behavior begins to take over is determined by $\kappa$, with larger values of $\kappa$ leading to an earlier change in regimes.

When the $e^{-8p}$ term begins to dominate, the amplitude of the oscillations will then decay as $e^{-2p}$. This is faster than its original $e^{-p}$ behavior and, more importantly, faster than the $e^{-4p/3}$ decay of the minimum of its effective potential. The oscillation amplitude will then decrease below the value of $\phimin$ and the field will fall into its minimum.

We have seen that the oscillatory motion of the chameleon in a quartic well creates large variations in the field's mass. Particle production must be considered, but the inclusion of these quantum effects does not result in the catastrophic effects experienced by other chameleon potentials. Instead the energy of excited modes is comparable to the temperature and the fraction of the initial energy lost to particle production is always less than 1 as long as $\kappa \lesssim 100$. Next we investigate whether quartic chameleons can also avoid the problems runaway models encounter when facing the kicks.

\section{Kicking the Quartic Chameleon}
\label{sec:kicks}

After the depletion of energy to particle production takes the chameleon field to or near the minimum of its effective potential, we can show that it will track its minimum until the onset of the kicks. The characteristic time scale for the evolution of the minimum is $\phimin/\mindot$, whereas the characteristic time scale of the response of the field is given by $m^{-1}$. When the field is in the minimum of its effective potential,
\beq
m^2 = \left.\frac{d^2V}{d\phi^2}\right|_{\phi=\phi_{\rm min}} = \frac{\kappa}{2}\phimin^2 = \left(\frac{9}{2}\kappa\right)^{1/3}\left(\frac{\beta\Sigma}{\Mpl}\rho_*\right)^{2/3}.
\eeq
If $m\gg \mindot/\phimin$, the field will adiabatically track its minimum. To compare these values, first we must determine $\mindot$:
\begin{align}
\mindot & = H \frac{d\phimin}{dp}, \nonumber \\
  & = \frac{2\beta}{\kappa\Mpl}\frac{H}{\phimin^2}\left(\rho_{*R}\frac{d\Sigma}{dT} + \Sigma\frac{d\rho_{*R}}{dT} \right)\frac{dT}{dp}, \nonumber \\
  & \simeq \frac{2\beta}{\kappa\Mpl}\frac{H}{\phimin^2}\rho_{*R}\left(\frac{d\Sigma}{dT} + 4\frac{\Sigma}{T}\right)(-T), \nonumber \\
  & = - \frac{1}{3}H\phimin\left(4 + \frac{T}{\Sigma}\frac{d\Sigma}{dT}\right),
\label{eq:mindot}
\end{align}
where we have used the definition of $\phimin$ given by Eq.~(\ref{eq:min}), the fact that $\rho_{*}\propto T^4 \propto a^{-4}$, and that $\phi\ll\Mpl$ implies $T\simeq T_J$. If $\Sigma$ is constant, the second term in parentheses drops out, and we recover the $\phimin \propto a^{-4/3}$ behavior we determined in Section II. Thus, when $\Sigma$ is constant, the evolution of the minimum is set by the expansion rate, and the characteristic time scale is approximately the Hubble time $H^{-1}$. Comparing this to the mass of the field in its minimum, we have
\begin{align}
\frac{m^2}{\left(\frac{\mindot}{\phimin}\right)^2} = \frac{4}{3}\frac{m^2}{H^2} & = \frac{4}{3}\left(\frac{9}{2}\kappa\right)^{1/3}\left(\frac{\beta\Sigma}{\Mpl}\rho_*\right)^{2/3} \left(\frac{3\Mpl^2}{\rho_*}\right),\nonumber\\
& = 4 \left(\frac{9}{2}\kappa\beta^2\right)^{1/3}\left(\frac{\Sigma^2\Mpl^4}{\frac{\pi^2}{30}g_*T^4}\right)^{1/3}.
\label{eq:m2H2}
\end{align}
We can see that, for reasonable values of $\kappa$ and $\beta$, the mass of the field at its potential minimum is much greater than $H$ for all $T \ll \Mpl$ while $\Sigma$ is constant.

Only if the field is positioned in a very small interval ($|\Delta\phi|\ll\phimin$) around $\phi=0$ is the effective mass less than $H$. However, even in this region, the constant nature of $\Sigma$ due to the QCD trace anomaly does not allow the field to become stuck due to Hubble friction. In this small region the effective potential is dominated by the matter coupling; if we neglect the driving term from the bare potential and use the fact that $V\ll\rho_{*R}$, we can simplify Eq.~(\ref{eq:fulleom}) to
\beq
\varphi^{\prime\prime} = -\varphi^\prime - 3\beta\Sigma.
\eeq
Integrating this equation for a chameleon initially at rest gives
\beq
\varphi^\prime\simeq3\Sigma\beta\left(1-e^{-p}\right).
\eeq
From this we see that $\varphi^\prime$ will increase toward a constant value until the field approaches its potential minimum and the bare potential can no longer be neglected. Thus, even if the chameleon begins at rest in a region where it has a low effective mass, Hubble friction will not prevent it from reaching its potential minimum. We have just shown in Eq.~(\ref{eq:m2H2}) that once it reaches this minimum it will then track it adiabatically.

Having established that the chameleon oscillates about its minimum at the onset of the kicks, we can now look at how they will affect the field's evolution. Comparing $m$ and $\mindot/\phimin$ when $\Sigma$ is no longer constant, we have
\begin{align}
\frac{m^2}{\left(\frac{\mindot}{\phimin}\right)^2} & = \frac{m^2}{H^2}{\frac{9}{\left(4 + \frac{T}{\Sigma}\frac{d\Sigma}{dT}\right)^2}}. \nonumber \\
\label{eq:mmindot}
\end{align}
The kick function $\Sigma$ displays two different types of behavior: at the beginning of the kicks when the temperature is greater than the mass of the particle species, $m_i$, $\Sigma\propto m_i^2/T^2$, and at the end of the kicks, when $T < m_i$, Boltzmann suppression makes $\Sigma\propto e^{- m_i/T}$. We can estimate the ratio in Eq.~(\ref{eq:mmindot}) using these two behaviors and the ratio $m^2/H^2$ given by Eq.~(\ref{eq:m2H2}). At the beginning of the kicks, when $d\Sigma/dT = -2\Sigma/T$,
\begin{align}
\frac{m^2}{\left(\frac{\mindot}{\phimin}\right)^2} & = \frac{m^2}{H^2}\frac{9}{4} \nonumber\\
 & \simeq \left(\frac{\Sigma\Mpl^2}{T^2}\right)^{2/3}.
\label{eq:tgm}
\end{align}
Even though $\Sigma \ll 1$, the quantity $(\Mpl/T)^2$ is more than sufficient to make this ratio $\gg 1$. At the end of the kicks, however, when $d\Sigma/dT = -m_i\Sigma/T^2$,
\begin{align}
\frac{m^2}{\left(\frac{\mindot}{\phimin}\right)^2} & = \frac{m^2}{H^2}\frac{9}{\left(4 + \frac{m_i}{T}\right)^2} \nonumber\\
 & \simeq \left(\frac{\Sigma T \Mpl^2}{m_i^3}\right)^{1/3},
\label{eq:mgt}
\end{align}
which is only $> 1$ as long as $\Mpl^2/m_i^3 > (\Sigma T)^{-1}$. This is true up until the end of the electron-positron kick, when the temperature and the value of $\Sigma$ continue to decrease past the point that $\Mpl^2/m_e^3$ can no longer compensate for their increasingly small values. For $\beta=0.1$ and $\kappa=2$, this occurs at approximately a temperature of 39 keV, when $\phimin \sim 10^{-33}\Mpl$.

\begin{figure}
\centering\includegraphics[width=3.4in]{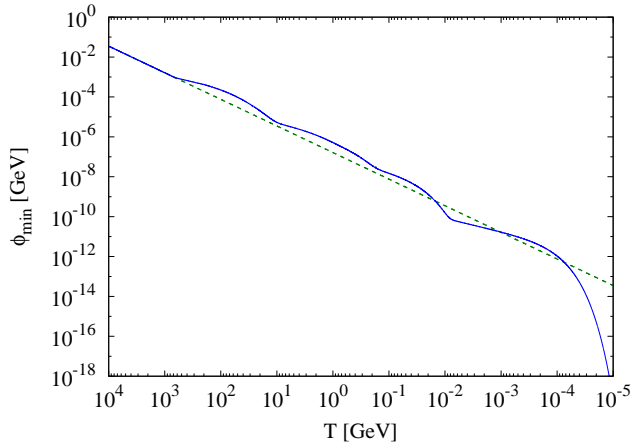}
\caption{The value of $\phimin$ during the kicks (blue, solid) and the extrapolation of the $e^{-4p/3}$ behavior experienced during the period when $\Sigma$ is constant (green, dashed). The value of $\phimin$ during the kicks actually decreases at a slower rate than it did when $\Sigma$ was constant until the end of the last kick. }
\label{sigma and min}
\end{figure}

\begin{figure}
\centering\includegraphics[width=3.4in]{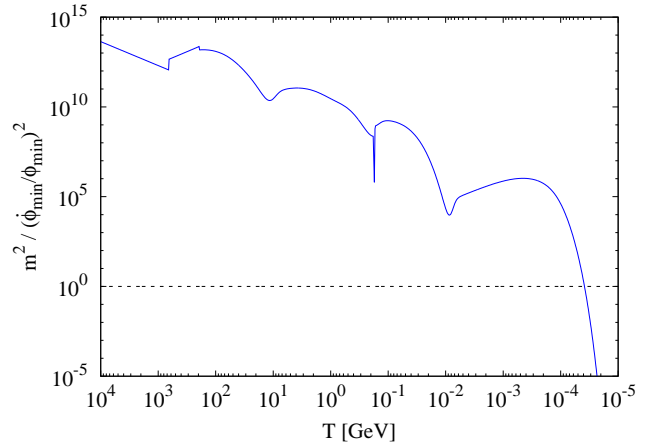}
\caption{The numerical evaluation of the ratio in  Eq. (\ref{eq:mmindot}), which is significantly greater than 1 throughout the kicks. It becomes less than 1 when $T\simeq3.9\times10^{-4}$GeV.}
\label{m2Sig2}
\end{figure}

Figure \ref{sigma and min} shows the value of $\phimin$ as a function of the temperature. The solid line shows the minimum under the influence of the kicks, while the dashed line shows the minimum following the $a^{-4/3}$ decay when $\Sigma$ is constant. The effect of the kicks for the most part is to slow the decrease in the minimum of the effective potential compared to this decay, until the very end of the kicks when Boltzmann suppression drastically decreases the value of $\Sigma$. In Figure \ref{m2Sig2} we have plotted the exact value of the ratio in Eq.~(\ref{eq:mmindot}), and we can see that the ratio is indeed much greater than 1 until after the last kick. Therefore the chameleon will track its minimum adiabatically until then.  When this occurs, even if we assume the field becomes entirely stuck while the potential minimum continues to decrease toward zero, the deviation of the field from its minimum cannot exceed the value at which it was stuck: $\sim\!\! 10^{-33}\Mpl$. This is clearly $\ll \Mpl$ and any implied variation in the particle masses would be completely negligible. When the Universe later becomes matter dominated and $\Sigma=1$, the field will once again be able to track the minimum of its effective potential.

\section{Conclusion}
\label{sec:end}

Since the chameleon model was first proposed as an alternative to dark energy \cite{KhouryLong, KhouryShort}, its cosmological impacts have been studied extensively. While it has been shown that chameleon theories cannot account for the expansion of the Universe without the addition of a constant term to its potential \cite{NoGo}, the field's sensitive dependence on its environment gives it remarkable properties that are of great interest. However, for most chameleon models, the same matter coupling that gives it its unique phenomenology leads these theories into trouble in the early Universe. The meV mass scale of runaway potentials is at odds with the GeV mass scale of SM particles, which accelerate the chameleon field to very high velocities when they become nonrelativistic. The hierarchy between these two energy scales leads to the quantum production of particles that radically alters the field's evolution. Without very weak couplings or highly tuned initial conditions these chameleon models cannot be trusted as effective field theories at the time of BBN \cite{AdrienneShort, AdrienneLong}.

In this paper, we have considered the quartic chameleon potential, which is not often studied in theories of chameleon gravity. A significant feature of this model is the fact that there is no mass scale in the potential: the chameleon's self-interaction is enough to ensure adequate screening. We have shown that this scale-free property of the potential allows the quartic chameleon to avoid the catastrophic effects of the small energy scales within runaway models.

After inflation, the quartic chameleon oscillates in its potential well. The amplitude of its oscillations are damped due to Hubble friction. In the classical treatment, the minimum of the field's effective potential decreases faster than the oscillation amplitude throughout radiation domination. Consequently, the field cannot reach its potential minimum before BBN, though the oscillation amplitude is always sufficiently small that the variation of the field from this minimum does not imply an unacceptable variation from known particle masses.

The rapid oscillations of the chameleon field cause changes in its effective mass that excite perturbations and lead to particle production. The effects of quantum particle production ensure that the field does reach its potential minimum while the Universe is radiation dominated. For large values ($\gtrsim 10$) of the self-interaction constant $\kappa$, the fractional loss of energy to these particles can be large, in which case the field loses all its energy in the course of a single oscillation. For smaller $\kappa$, the energy lost to particles constitutes only a small fraction of the field's energy. This much slower energy loss accumulates over multiple oscillations and introduces an additional decay term to the oscillation amplitude which allows the field to catch its minimum after many oscillations. At this point, the field will adiabatically track its potential minimum. It will track this minimum until the very tail end of the kicks, when the Boltzmann suppression of $\Sigma$ decreases the value of $\phimin$ faster than the field can follow. The value of the field at this point is sufficiently small that any deviation of particle masses implied by the deviation of the field from its potential minimum are entirely negligible.

The energy of the modes that are excited in the quartic model are on order of the temperature: highly energetic modes are only excited at high temperatures. This is an important contrast to runaway models, which experience extremely energetic fluctuations at relatively low temperatures and can no longer be treated as EFTs during BBN. While quantum corrections lead to extremely energetic fluctuations in the field and a breakdown in calculability for runaway models, quantum corrections to the quartic potential are comparatively small and are, in fact, necessary to ensure the field can reach its minimum. Once it reaches the minimum of its effective potential the chameleon can then adiabatically track this minimum even throughout the kicks. Thus, the quartic chameleon's scale-free nature means is not susceptible to problems arising from a hierarchy of scales and can remain a well-behaved effective field theory throughout the evolution of the early Universe.

\acknowledgements
We thank Kayla Redmond for her comments on our manuscript. C.M. acknowledges support from the Bahnson Fund at the University of North Carolina at Chapel Hill.

\appendix
\setcounter{secnumdepth}{0}
\section{Appendix: Energy Densities in Chameleon Gravity}
\label{sec:Appendix}

In this appendix we take a closer look at the energy evolution of different quantities in the Jordan and Einstein frames.

\subsection{1. Energy Density of Matter and Radiation}
\label{sec:energydensity}

In order to consider the quantities that are conserved in both the Einstein and Jordan frames recall how the energy density and scale factor in each frame are related, namely, $\tilde{\rho} = e^{4\beta\varphi} \rho_*$ and $\tilde{a} = e^{-\beta\varphi}a_*$, where we have used the dimensionless variable $\varphi = \phi / \Mpl$ . Matter and the chameleon field do not interact in the Jordan frame and so the matter stress-energy tensor is conserved, $\tilde{\nabla}_\mu \tilde{T}^\mu{}_\nu = 0$, and we can write the conservation equation in the Jordan frame,
\beq
\tilde{\rho} \propto \tilde{a}^{-3(w+1)}.
\eeq

Exchanging the Jordan frame quantities for those of the Einstein frame, we have
\begin{align}
\rho_* e^{4\beta\varphi} & \propto \left(a_*e^{-\beta\varphi}\right)^{-3(w+1)},\nonumber\\
\rho_* e^{4\beta\varphi} & \propto a_*^{-3(w+1)}e^{3\beta\varphi(w+1)},\nonumber\\
\rho_* e^{\beta\varphi(1-3w)} & \propto a_*^{-3(w+1)}.
\label{eq:energyconservation}
\end{align}
For matter we have $w=0$ and Eq.~(\ref{eq:energyconservation}) becomes
\begin{align}
\rho_{*m} e^{\beta\varphi} & \propto a_*^{-3}.
\label{eq:matterconservation}
\end{align}
Clearly, the Einstein-frame energy density in matter, $\rho_{*m}$, does not scale as $a_*^{-3}$.  This follows from our earlier statement in Section II that the stress-energy tensor that is conserved in the Einstein frame is not ${T_*^\mu}_\nu$, but the sum of ${T_*^\mu}_\nu$ and the stress-energy tensor of the chameleon field. Often it has been the practice to define the left-hand side Eq.~(\ref{eq:matterconservation}) as the matter density as it is the quantity that follows the conservation equation in the Einstein frame \cite{KhouryLong, Brax}.

For radiation, however, $w=1/3$ and we find from Eq.~(\ref{eq:energyconservation}) that
\begin{align}
\rho_{*R} & \propto a_*^{-4}.
\label{eq:raditationconservation}
\end{align}
In both frames the energy density in radiation is proportional to $a_*^{-4}$, and it follows that $H_* \propto a_*^{-2}$.

\subsection{2. Energy Density of the Chameleon}
\label{sec:pressure}

In Section II we used the canonical definitions of the energy density and pressure of a scalar field, namely \mbox{$\rho_\phi = \phidot^2/2 + V$} and $P_\phi = \phidot^2/2 - V$, to determine its equation of state, $w$. However, when defining $w$ in this way, because the field's coupling to matter allows it to exchange energy with the matter fields, the chameleon does not follow the conservation equation,
\beq
\frac{\rhodot_\phi}{\rho_\phi} - 3 H_*^2 (1+w) \neq 0.
\label{eq:conservation}
\eeq
If we instead introduce a new pressure,
\beq
P_n = \frac{1}{2}\phidot^2 - V - \frac{1}{3 H_*}\phidot \frac{\beta}{\Mpl}\Sigma\rho_*
\eeq
we can see that with this new definition, the field now obeys the conservation equation,
\begin{align}
& \rhodot_\phi + 3 H_* (\rho_\phi + P_n) = \nonumber \\
& = \phidot \ddot{\phi} + \frac{dV}{d\phi}\phidot + 3 H_* \left(\frac{\phidot^2}{2} + V + \frac{\phidot^2}{2} - V - \frac{1}{3 H_*}\phidot \frac{\beta}{\Mpl}\Sigma\rho_*\right) \nonumber \\
& = \phidot\left(\ddot{\phi} + 3H_*\phidot + \frac{dV}{d\phi} - \frac{\beta}{\Mpl}\Sigma \rho_*\right).
\label{eq:conserved}
\end{align}
The terms in parentheses in the last line make up the chameleon's equation of motion, Eq.~(\ref{eq:eom}), and the entire quantity in parentheses is indeed equal to 0. While we used the canonical form of the pressure in the text, we also took its average value over many oscillations, and the term which we have added to the potential is not positive-definite, and will average to 0.

Not only will it average to 0, but we can also show that the additional contribution to the new pressure is negligible compared to the usual terms. Using the relation $\phidot = H_*\Mpl\phiprime$, we can rewrite $P_n$ as
\begin{align}
P_n & = \frac{1}{2}(H_*\Mpl\phiprime)^2 - V - \frac{1}{3}\phiprime\beta\Sigma\rho_* \nonumber\\
    & = \frac{1}{6}\rho_*\varphi'^2 - V - \frac{1}{3}\phiprime\beta\Sigma\rho_*.
\end{align}

The maximum values of the first two terms, the kinetic and potential energies of the field, are equal and so to compare the relative contribution of the last term, we will look specifically at how it compares the kinetic energy. The maximum value reached by $\phiprime$, given in Eq.~(\ref{eq:phiprimemax}), can be broken down further using the initial values of the field and $H_*$,
\begin{align}
\phiprime_\mathrm{max} & = \sqrt{\frac{\kappa}{12}}\sqrt{\frac{3}{\rho_{*,i}}}\left(\frac{6\beta \Sigma_i\rho_{*,i}}{\kappa\Mpl}\right)^{2/3} \nonumber\\
& \simeq \left(\frac{\beta^4}{\kappa}\right)^{1/6}\left(\frac{81}{4}\Sigma_i^4\frac{\rho_{*R,i}}{\Mpl^4}\right)^{1/6}.
\end{align}
Combining numerical factors and using $T_{J,i} = 10^{16}$ GeV we have
\beq
\phiprime_\mathrm{max} = 0.19 \left(\frac{\beta^4}{\kappa}\right)^{1/6}.
\eeq
The relative contribution of the two terms is then $\beta\Sigma/\phiprime_\mathrm{max} \sim 0.005(\kappa\beta^2)^{1/6}$, which for even some of the larger values of $\kappa$ and $\beta$ allowed ($\kappa = 100$, $\beta = 1$) is still much less than 1. Therefore, the additional term is a negligible contribution to the canonical pressure, and our use of the conservation equation, Eq.~(\ref{eq:virialenergy}), is valid.

\end{document}